# AN ABSTRACT OF THE THESIS OF

Robert F. Melendy for the degree of Master of Science in Electrical and Computer Engineering presented on December 18, 1997. Title: Bang-Bang Control Development of Permeability Changes in a Membrane Model

Abstract approved: ______________________________

Ronald R. Mohler

The application of systems and control theory to membrane physiology is presented here. Modeling efforts have focused on describing those physiologically realistic mechanisms which govern the regulation of membrane permeability in nerve. The motivation behind identifying such mechanisms lies in understanding the morphology of neural activity on a meaningful and analytically tractable level. The suggested merit of integrating control theory into the analysis lies in providing how a membrane effectively adapts to changes in permeability and through what governing mechanisms. The value in producing such an understanding lies in mirroring biological reality in a more formal manner than could be achieved solely through experimental means. A bang-bang control policy describing the permeability correction mechanisms is developed using Liapunov's Stability Criteria. Both changes in membrane potential and kinetic rates are required to implement the policy. The policy describes the inherent mechanisms of the membrane which act to drive its permeability from unstable firing to the resting potential state. It is shown that these permeability changes in state are governed by a switching function that depends on the membrane potential and a dominant controlling parameter. The control policy is discussed in the context of solutions of the Hodgkin-Huxley Equations of Ionic Hypothesis.

# Bang-Bang Control Development of Permeability Changes in a Membrane Model

by

Robert F. Melendy

A THESIS

submitted to

Oregon State University

in partial fulfillment of
the requirements for the
degree of

Master of Science

# TABLE OF CONTENTS

## TABLE OF CONTENTS (Continued)

# LIST OF FIGURES

# Bang-Bang Control Development of Permeability Changes in a Membrane Model

## Chapter 1. Thesis Overview

### 1.1. Introduction: Thesis Objective and Motivation

#### 1.1.1. Research Objective

The objective of this thesis is to apply rigorous engineering methods in an effort to model physiologically realistic mechanisms which govern the regulation of membrane permeability in nerve. The proposed model development will be discussed in the context of the Hodgkin and Huxley Equations of Ionic Hypothesis [H2] (henceforth referred to as the Hodgkin-Huxley Equations). It is suggested that systems analysis and control theory may play a role in describing the ways in which the Hodgkin-Huxley membrane adapts to changes in permeability from rest. Based on activation and response studies [B3, D1], the conjecture is that an underlying mechanism may exist which acts to stabilize membrane permeability through a switching or bang-bang type action. The underlying object in performing these analyses is to provide a formal development and suggested implementation of a so-called "membrane control policy."

#### 1.1.2. Motivation and Value

The motivation behind identifying the switching and control mechanisms inherent in the Hodgkin-Huxley membrane model [H2] lies in understanding the morphology of neural activity on a meaningful and analytically tractable level. The value in producing such an understanding lies in mirroring biological reality in a more formal manner than could be achieved solely through experimental means [B1, G1]. For the experimentalist, say, the insight gained from a formal model description may provide guidance in developing advanced experiments. In turn, the observations gained from such experiments could provide insight which sustains the systems analyst in generalizing the model for relevant applications (such as in advanced computer architectures and control systems).

The suggested merit of integrating control theory into the analysis lies in describing the ways in which a membrane effectively adapts to changes in permeability and through what governing mechanisms. The value in this is two-fold: First, previously derived models developed through physiological hypotheses and experimental means may be rigorously challenged and perhaps verified. Secondly, providence of a formal membrane regulation model could possibly lend to clinical, engineering, and pharmacokinetic developments.

The motivation, say, for pharmacological laboratory research may be that of introducing an additive control mechanism through drug dosage administration. The research pharmacist may experimentally induce drugs into a living organism and observe what cellular responses take place [B2, W1]. In contrast, a systems approach may entail the rigorous development of an equivalent cellular model. The systems model may lend a deeper level of understanding of the behavioral mechanisms inherent in the cell's structure which predominantly respond to the drug disturbance. Clinical developments from such a formal description could possibly lead to the improved regulation of drug infusion rates as administered through a controlled drug-delivery system.

To incite motivation in an electrical engineering application, one might consider recent developments of artificial neural networks (ANN's) [A1, C2].

Typical ANN's are obscurely similar to the perplexing computational structure of a real neurological system [M6, M8]. Therefore, most neural network structures are based on a model of a *neuron* that captures the most basic features of the real system. The design of an ANN typically includes some specification of the real neurological system, the architecture, and the learning process. One might find value in a formal systems description of membrane permeability adaptation in nerve. Understanding which mechanisms govern the way in which a membrane obtains a stabilizing permeable state in the presence of a cellular disturbance could possibly suggest a means for minimizing output error over a training set.

## 1.2. The Role of Recent Developments in Neurogenic Dynamic Modeling

### 1.2.1. History in Brief

In 1939, Cole, Curtis, Hodgkin and Huxley succeeded in directly measuring the membrane potential from within the nerve fiber of a squid axon [N1]. From their intracellular recordings, they developed a mathematical model based on the electrical properties of the squid's nerve membrane. The model supporting their theory are the so-called Equations of Ionic Hypothesis [H2]. This theory was further used to infer an electric circuit equivalent emulating membrane capacitance charging and ionic movement within the membrane (see Chapter 2).

### 1.2.2. Related Literature

In an effort to make contributing developments to the work originally developed by Hodgkin and Huxley [H2], researchers from various areas of biophysics, physiology, and applied mathematics have reported on developments of various dynamical foundations [H3]. Modern analyses have focused primarily on subthreshold oscillatory characteristics and repetitive firing in nerve during membrane permeability changes [R2, S2]. As Hodgkin and Huxley have shown in their classic paper [H2], permeability changes such as these come about through a propagated action potential (as triggered by a stimulus current applied to the membrane).

Time-and-again, the focus of current researcher's developments have brought to bear that permeability changes in squid (and some molluscan) giant neurons model after the behavior of a relaxation oscillator network [S3] (i.e., differential describing models which contain a fast and a slow mode — see Chapter 2, §2.2). It is based on this working hypothesis that much current research in dynamical membrane modeling is being performed.

In performing the literature search, there appeared to be a sparsity of a formal systems analysis which investigates how membrane permeability is regulated in the presence of an action potential. The search included papers where a formal development of a "membrane control policy" might be found.

Pinsky *et al* [P1] (1996) have been investigating bifurcations of membrane models governed by differential equations with fast and slow variables (as in models for bursting oscillations). In performing this analysis, Pinsky has considered first the slow variables as parameters and proceeds to study the fast modes. Pinsky claims to have proved, analytically, certain relations about the bifurcations of the fast mode and those of the full system. This is more-or-less of a similarity study, however, meaning that a comparison of bifurcation types among the fast and full system is made. To this end, no descriptive policy of an underlying correction mechanism of membrane permeability is provided.

In reference to membrane permeability adaptation, Mejia and Lynch [M1]

(1995) have recently described the effect of changing transport properties in the concentration profiles of ionic compounds within a single cell. They describe this phenomenon using a reaction-diffusion model. Theirs is a bioelectric field-type problem, however, and is not formulated for a state-analysis.

Sherman *et al* [S4] (1995) have developed a topological classification scheme for membrane bursting oscillations. Their methodology is to compute a two-parameter plane of all possible dynamic behaviors of simple (two-variable) fast mode generating systems. The claim is that when combined with the slow variables, these constitute all possible oscillators within a certain family. Indeed, Sherman has formulated a mathematically elegant foundation. However, this model only classifies bursting patterns and provides no description of how ionic switching mechanisms act to stabilize membrane permeability.

The control of vasomotion in nerve has been observed by Gonzalez-Fernadez [G3] (1994). The crux of this work lies in observing *in vitro* depolarization of a cell membrane and the resulting promotion of stabilizing ionic entry. Yet, this work is purely experimental and is not supported by a dynamical analysis. From a systems viewpoint, the cell depolarization emulates a metabolic disturbance wherein the correction mechanism of the hyperpolarizing control takes the form of the stabilizing ionic entry. A natural follow-up of this research would be the identification of the ionic correction mechanisms which govern vasomotion stabilization (similar to the mechanism studied here).

Li, Stojilkovic, and Keizer [L1] (1994) have developed a model yielding a good, quantitative description of membrane electrical activity and its link to the accompanying oscillations in calcium ions. Specifically, their model is used to describe calcium response release mechanisms in gonadotrophins in the presence of a hormonal disturbance. From a systems viewpoint, their model does not identify which mechanisms regulate these calcium concentrations.

Interestingly, Rall *et al* [R1] (1993) have developed a theoretical foundation of dendritic function. Rall has demonstrated that under certain conditions, bistable oscillatory behavior has been observed in vertebrate motoneurons. From a systems viewpoint, Rall's work motivates the development of a bilinear control policy. A suggested approach might entail the development of a so-called sliding control [M3, M5]. The function of this control would hold the hyperplane along the regions of which admissible bistable oscillatory states are separated.

Golomb, Rinzel, and Wang [G2] (1994) have recently developed a model for describing the genesis of sleep spindle oscillations (otherwise classified as *rhythmogenesis in thalamic networks*). They address the so-called *isolated reticular thalamic nucleus* as a generator of spindle oscillations. Right or wrong, their study goes as far as to conclude that the bursting of each model neuron is at a Hopf bifurcation.

Pinsky [P2] (1994) has explored synchronism and clustering in hippocampal networks. In this study, Pinsky has formulated a series of simplified models of hippocampal neurons as neural networks. Pinsky identified certain relaxation properties inherent in the model, making the conjecture that bursting oscillations are robust in the face of considerable cell disturbances. No link between the relaxation modes nor the underlying correction mechanisms between bursts was identified however.

## 1.3. Overview of Proposed Analysis, Expectations, and Order of Presentation

In studying the literature and the Equations of Ionic Hypothesis [H2], it is understood that changes in a membrane's potential — and consequently, ionic concentrations which regulate the permeability of the membrane — are the governing mechanisms one must account for if a formal control policy is to be developed. It is suggested therefore that a control policy be developed so as to describe membrane permeability adaptation in a squid giant axon in the face of a propagated action potential.

It is suggested that this analysis take place for subthreshold oscillations so as to examine the membrane return-to-rest after release of an external disturbance [M7, R2]. Furthermore, switching mechanisms inherent in the control policy should indicate the dominant mechanisms which govern how the membrane obtains a reachable, stabilizing permeable state.

The foregoing developments will be performed in the context of the original Hodgkin-Huxley equations (as derived in their 1952 paper [H2]). It is important to note at the outset that these equations are in fact, analytically intractable, and so solutions may only be obtained through a numerical integration. The author therefore suggests converting these equations into state form, thereupon integrating them using a digital computer. In doing so, the integrations will allow the system to be observed in its complete nonlinear and time-varying form. This work is presented in detail in Chapters 2 and 3.

The reason for the state formulation is two-fold, however, as presentation of the Hodgkin-Huxley equations in this manner lead naturally to the development of a bang-bang control policy. The details of this policy and the methods used to develop it are the subject of Chapter 4. Conclusions based on the controller's characteristics, the underlying switching mechanisms which act to stabilize membrane permeability, and a suggested implementation of the "membrane control policy" will also be presented in that chapter.

# Chapter 2. The Equations of Ionic Hypothesis

## 2.1. The Hodgkin-Huxley Model

The equations which describe the kinetic properties of ionic conductances in a squid's membrane are presented here. Hodgkin and Huxley hypothesized that the net current $I_M$ which flows into a unit area of membrane surface is the sum of the current $I_C$ flowing into the membrane capacitance $C_M$ (per unit area) and the ionic current $I_i$ associated primarily with sodium and potassium species:

$$I_M = m^3 h \bar{g}_{Na}(E - E_{Na}) + n^4 \bar{g}_K(E - E_K) + \bar{g}_L(E - E_L) + C_M \frac{dE}{dt} \qquad (2.1\text{-}1)$$

$$\frac{dm}{dt} = \alpha_m(1 - m) - \beta_m m \qquad (2.1\text{-}2)$$

$$\frac{dh}{dt} = \alpha_h(1 - h) - \beta_h h \qquad (2.1\text{-}3)$$

$$\frac{dn}{dt} = \alpha_n(1 - n) - \beta_n n \qquad (2.1\text{-}4)$$

where $E$ is the membrane potential (in $mV$). Likewise, the potential of the sodium, potassium, and leakage channels are denoted by $E_{Na}$, $E_K$, and $E_L$, respectively. The maximum conductance associated with each species is denoted by $\bar{g}_{Na}$, $g_K$, and $g_L$ (in units of $mmhos/cm^2$). The so-called controlling parameters $m$, $h$, and $n$ are time-varying coefficients $\in (0, 1)$ and represent the probability that any particular channel (or *pore*) is open to the flow of ionic currents $I_{Na}$ and $I_K$. Specifically, $m$ and $h$ are associated with two types of sodium channels, whereas $n$ is associated solely with potassium.

In reference to differential equations (2.1-2) through (2.1-4), each $\alpha$ and $\beta$ represents an experimentally observed rate-constant (derived from kinetic theory) and is approximated by a smooth, mathematical function of the membrane voltage $E$. Thus, $\alpha = \alpha(E)$ and $\beta = \beta(E)$. These functions for the squid axon at a temperature of 6.3 °C are [H1]

$$\alpha_m = \frac{0.1(E + 40)}{1 - e^{-(E + 40)/10}} \tag{2.1-5}$$

$$\beta_m = 0.108e^{-(E/18)} \tag{2.1-6}$$

$$\alpha_h = 0.0027e^{-(E/20)} \tag{2.1-7}$$

$$\beta_h = \frac{1}{1 + e^{-(E + 35)/10}} \tag{2.1-8}$$

$$\alpha_n = \frac{0.01(E + 55)}{1 - e^{-(E + 55)/10}} \tag{2.1-9}$$

$$\beta_n = 0.0555e^{-(E/80)} \tag{2.1-10}$$

## 2.2. Membrane Electrical Equivalent and Engineering Circuit Approximation

Equations (2.1-1) through (2.1-10), along with the required initial conditions and parameter values, form a sufficient set of relations necessary in solving for the membrane potential $E$ and the controlling parameters $m$, $h$, and $n$. The next task will be to numerically integrate equations (2.1-1) through (2.1-4), thus solving for these state variables. As a consequence, intracellular currents and membrane changes during a propagated action potential may be fully simulated.

The integrated results will allow for the computation of the ionic currents $I_{Na}$, $I_K$, and the leakage term $I_L$, in addition to the *time-varying* conductances $g_{Na}$, $g_K$, and $g_L$. Identifying each ionic current in equation (2.1-1), it is evident that

$$I_{Na} = m^3 h \bar{g}_{Na}(E - E_{Na}) \tag{2.2-1}$$

$$I_K = n^4 \bar{g}_K (E - E_K) \tag{2.2-2}$$

from which the time-varying conductances of sodium and potassium are [H1]

$$g_{Na} = \frac{I_{Na}}{(E - E_{Na})} \tag{2.2-3}$$

$$g_K = \frac{I_K}{(E - E_K)} \tag{2.2-4}$$

Figure 2.2.1 illustrates the membrane electrical equivalent of the squid axon [N1].

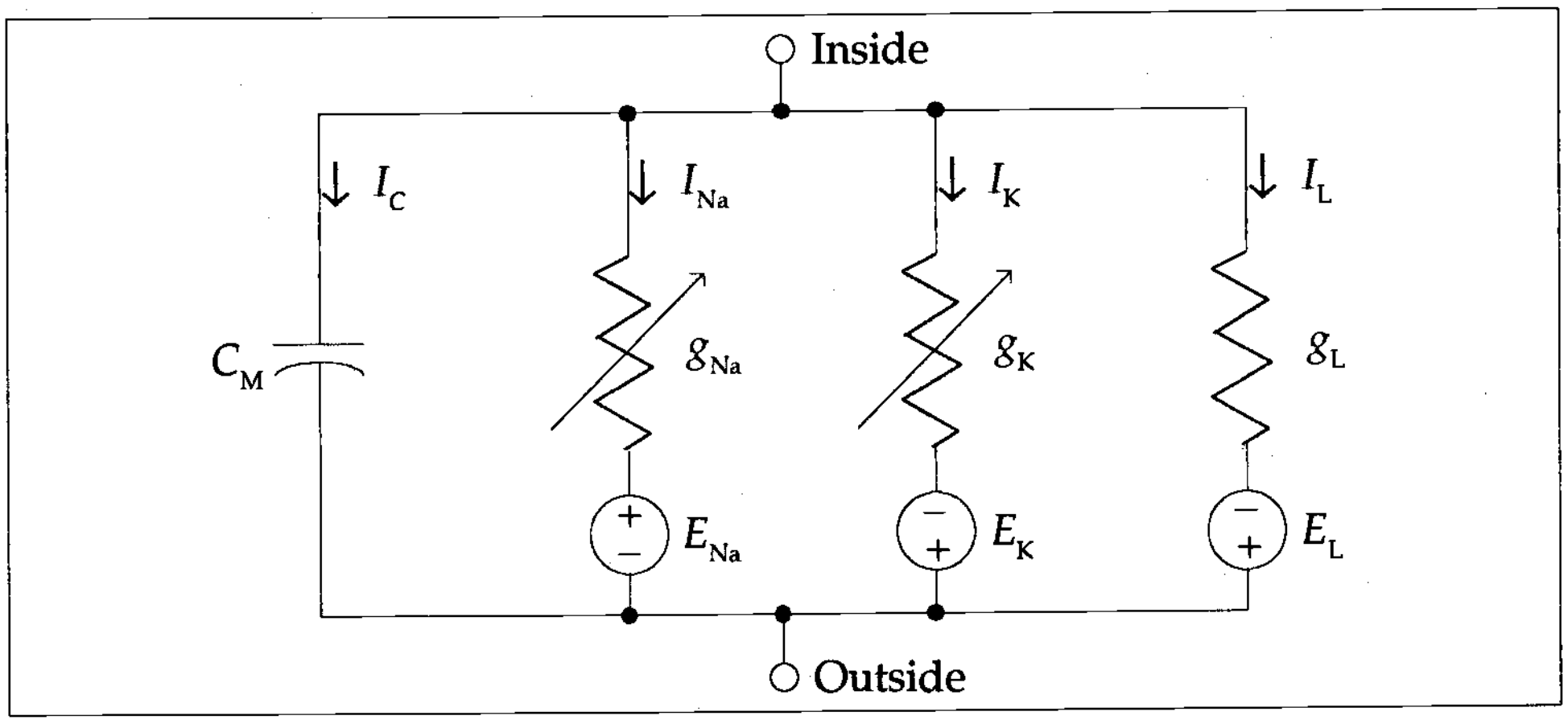

Figure 2.2.1. Electrical equivalent circuit for the membrane of the squid axon.

In reference to Chapter 1, §1.2.2, it was implied that post Hodgkin-Huxley research efforts [S3] have confirmed that permeability changes in the squid membrane behave much like a relaxation oscillator network (i.e., a parametrically controlled linear system with nonlinear feedback). Elaborating on this hypothesis from an electrical engineering perspective, the membrane electrical equivalent of Figure 2.2.1 might then be modeled, approximately, as a free running pulse-type circuit [M2]. A generic pulse circuit structure typically contains energy storage elements and a nonlinear active device as seen in Figure 2.2.2.

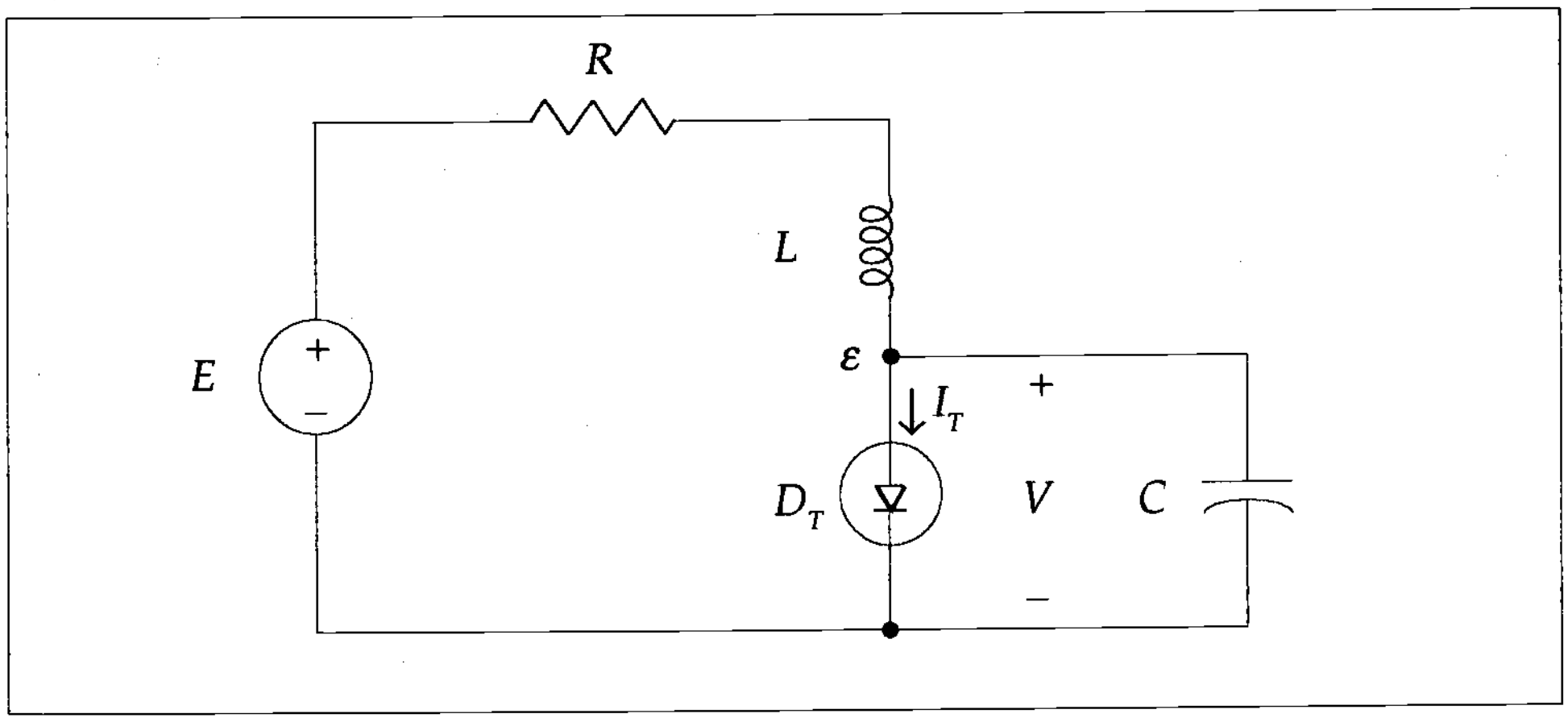

Figure 2.2.2. Engineering circuit approximation of the Hodgkin-Huxley membrane.

The circuit model of Figure 2.2.2 is described by the second-order nonlinear system

$$L\dot{i} = E - V - iR$$

$$C\dot{V} = i - g(V) \qquad (2.2\text{-}5)$$

also known as the *Bonhoeffer-Van der Pol equations* [V1]. The first expression in (2.2-5) comes about from the application of Kirchoff's voltage law around the outer loop. The second equation results from applying Kirchoff's current law to node $\varepsilon$. The term $g(V)$ in this second expression describes the voltage-controlled nonlinear conductance of the tunnel diode $D_T$. This device is a semiconductor junction diode which is heavily doped and which has an extremely narrow junction thickness [M2]). Consequently, this device has a negative resistance region, an extremely high switching-speed, and is relatively free from the effect of temperature.

In the generic sense, the Bonhoeffer-Van der Pol system of (2.2-5) approximates the excitability and threshold phenomena of the Hodgkin-Huxley nerve membrane in a way that is mathematically tractable. There is one identifying discrepancy between the behavior of the Bonhoeffer-Van der Pol system and that of the membrane electrical equivalent: Whereas the network of Figure 2.2.2 can sustain repetitive firing (i.e., it is a free running oscillatory network), such activity has not been consistently observed in experiments on real squid axon [B3, C1, M7, S1].

Due to the analytical intractability of the Hodgkin-Huxley system, membrane firing and permeability phenomena may only be obtained through numerically integrating (2.1-1) through (2.1-4). Setting up (2.1-1) through (2.1-4) for a numerical integration, followed by simulation results of pertinent ionic parameters are the topics of Chapter 3.

# Chapter 3. State Formulation of the Hodgkin-Huxley Equations and the Presentation of Numerical Results

## 3.1. State Structuring of the Hodgkin-Huxley Equations

In Chapter 2 it was stated that equations (2.1-1) through (2.1-10), along with required initial conditions and parameter values, form a necessary and sufficient set of relations necessary in solving for the membrane potential $E$ and the controlling parameters $m$, $h$, and $n$. With these solutions, the ionic currents and the time-varying conductances of equations (2.2-1) through (2.2-4) may be obtained. Before proceeding to integrate equations (2.1-1) through (2.1-4), it will prove useful to express these relations in state form (not only for a numerical simulation, but for the dynamical analysis of Chapter 4).

The first step is to identify all state variables, specifically, $E$, $m$, $h$, and $n$. Let $E = x_1$, $m = x_2$, $h = x_3$, $n = x_4$ and make these substitutions into equations (2.1-1) through (2.1-4). Then, isolate the capacitive current term $C_M\dot{x}_1$ to one side of equation (2.1-1) and divide through by the membrane capacitance $C_M$ such that

$$\dot{x}_1 = \frac{I_s(t) - x_2^3 x_3 \bar{g}_{Na}(x_1 - E_{Na}) + x_4^4 \bar{g}_K(x_1 - E_K) + \bar{g}_L(x_1 - E_L)}{C_M} \tag{3.1-1}$$

$$\dot{x}_2 = \alpha_m(1 - x_2) - \beta_m x_2 \tag{3.1-2}$$

$$\dot{x}_3 = \alpha_h(1 - x_3) - \beta_h x_3 \tag{3.1-3}$$

$$\dot{x}_4 = \alpha_n(1 - x_4) - \beta_n x_4 \tag{3.1-4}$$

For generality, the membrane current $I_M$ has been set to a stimulus current $I_s(t)$.

Upon expanding and reordering terms in equation (3.1-1)

$$\dot{x}_1 = \frac{I_s(t) - \bar{g}_{Na} x_1 x_2^3 x_3 - \bar{g}_{Na} E_{Na} x_2^3 x_3 + \bar{g}_K x_1 x_4^4 - \bar{g}_K E_K x_4^4 + \bar{g}_L x_1 - \bar{g}_L E_L}{C_M}$$

In dividing each term through by $C_M$, let the coefficients $I_s(t)/C_M = k_s$, $\bar{g}_{Na}/C_M = k_1$, $\bar{g}_{Na}E_{Na}/C_M = k_2$, $\bar{g}_K/C_M = k_3$, $\bar{g}_K E_K/C_M = k_4$, $\bar{g}_L/C_M = k_5$, and $\bar{g}_L E_L/C_M = k_6$. Then, upon grouping coefficients of like terms in $x$

$$\dot{x}_1 = k_s + (k_2 - k_1 x_1) x_2^3 x_3 + (k_4 - k_3 x_1) x_4^4 - k_5 x_1 + k_6 \tag{3.1-5a}$$

Now it is desired to place equations (3.1-2) through (3.1-4) in a form similar to (3.1-5*a*). First, factor (3.1-2) through (3.1-4) to obtain

$$\dot{x}_2 = \alpha_m - (\alpha_m + \beta_m) x_2$$

$$\dot{x}_3 = \alpha_h - (\alpha_h + \beta_h) x_3$$

$$\dot{x}_4 = \alpha_n - (\alpha_n + \beta_n) x_4$$

Then, upon substitution of each $\alpha$ and $\beta$ expression (2.1-5) through (2.1-10) into the above, the state variable $x_1$ is consequently incorporated. Each rate-constant may hence be expressed in the form $\alpha = \alpha(x_1)$ and $\beta = \beta(x_1)$. Next, let $\alpha_m = f_1(x_1)$, $(\alpha_m + \beta_m) = f_2(x_1)$, $\alpha_h = f_3(x_1)$, $(\alpha_h + \beta_h) = f_4(x_1)$, $\alpha_n = f_5(x_1)$, and $(\alpha_n + \beta_n) = f_6(x_1)$. Thus

$$\dot{x}_2 = f_1(x_1) - f_2(x_1) x_2 \tag{3.1-5b}$$

$$\dot{x}_3 = f_3(x_1) - f_4(x_1) x_3 \tag{3.1-5c}$$

$$\dot{x}_4 = f_5(x_1) - f_6(x_1)x_4 \qquad (3.1\text{-}5d)$$

Equations (3.1-5*a*) through (3.1-5*d*) are now set up for a numerical integration, the results of which are illustrated in Figure 3.1.1. These integrations were performed using a fourth-order Runge-Kutta algorithm. This, along with the required initial conditions and parameter values are listed in the appendix.

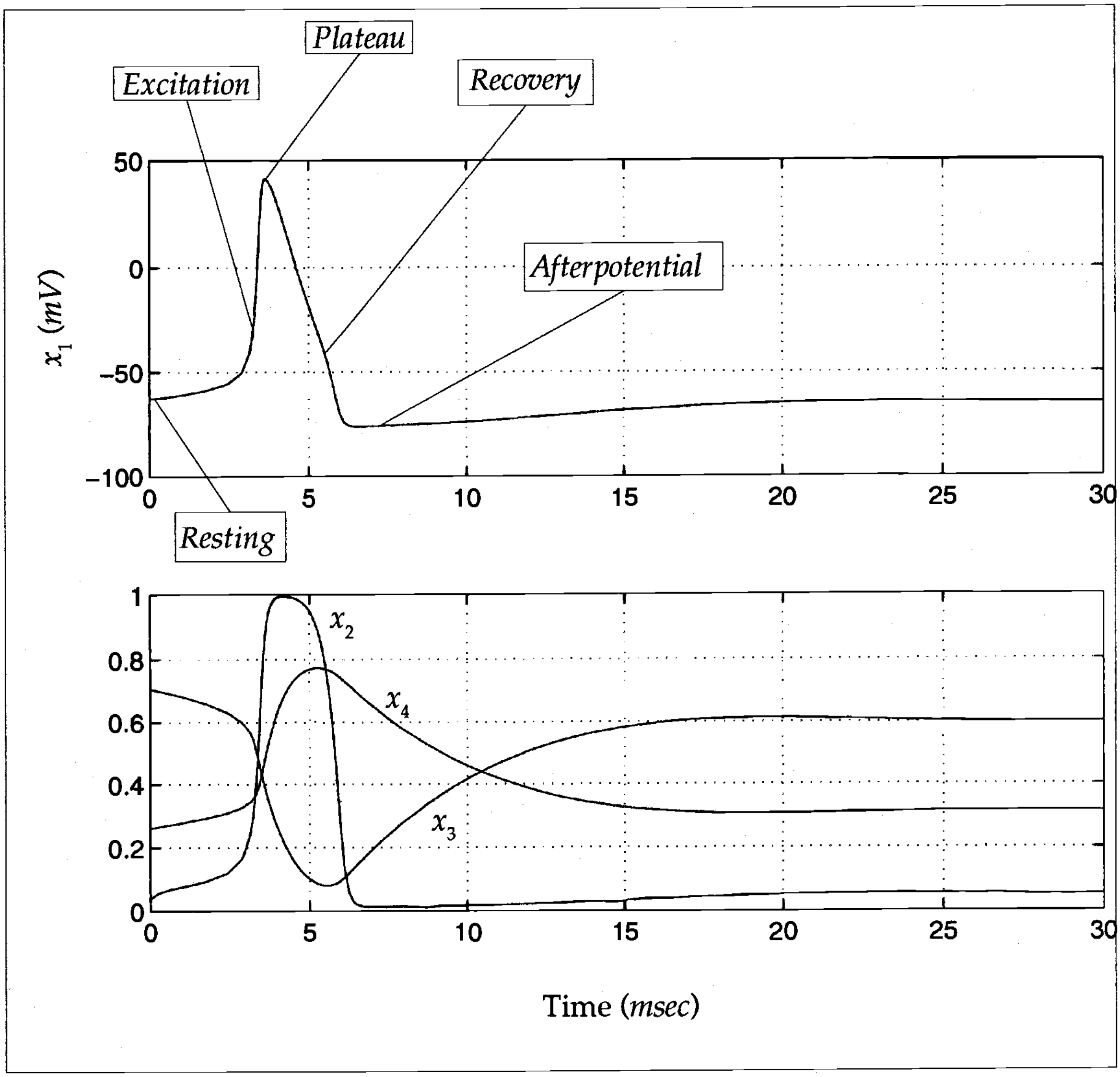

Figure 3.1.1. Membrane potential and controlling parameters at $T$ = 6.3 °C.

### 3.2. Membrane Currents and Time-Varying Conductances

An illustration of the current changes which take place in the membrane is shown in Figure 3.2.1. These results are computed using Equations (2.2-1) and (2.2-2). Potassium and sodium current flows during a propagated action potential are illustrated in (*a*). Plot (*b*) shows the membrane capacitive current flow.

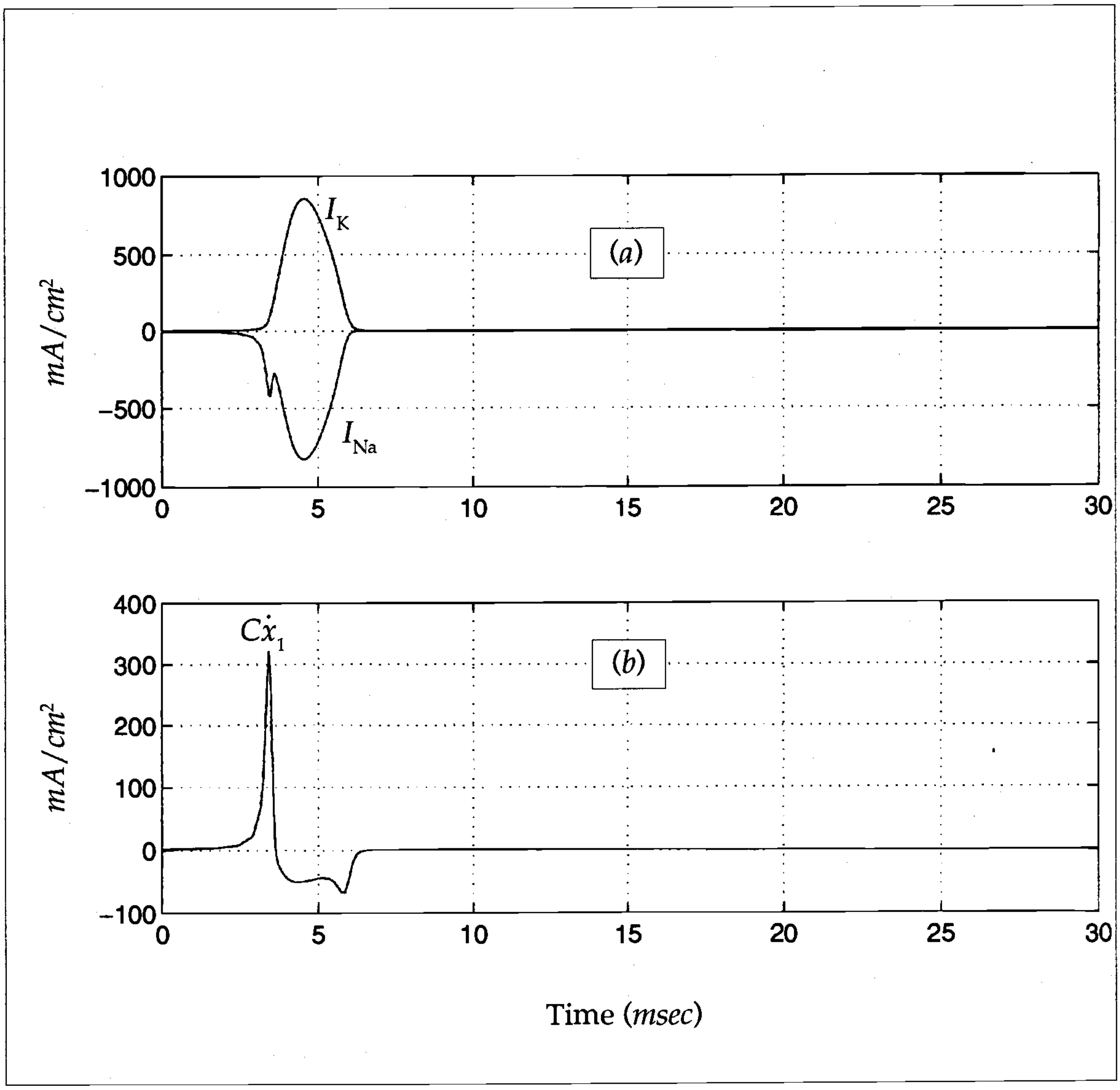

Figure 3.2.1. Membrane currents: (*a*) Potassium, sodium, and (*b*) capacitive.

An illustration of the conductance changes which take place in the membrane is provided in Figure 3.2.2.

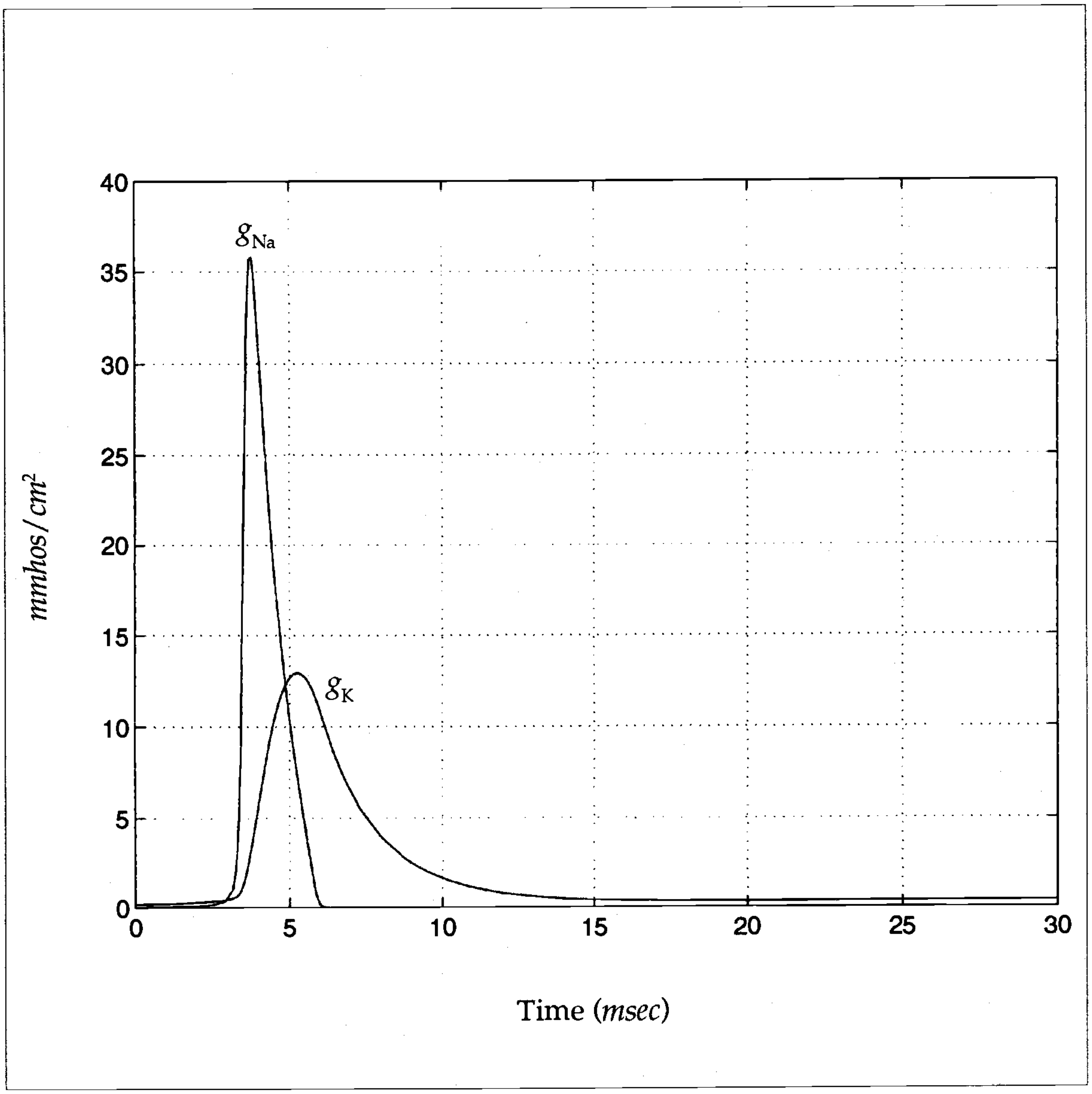

Figure 3.2.2. Potassium and sodium time-varying membrane conductances.

Clearly, the potassium conductance portrays the slow mode of the differential system. This observation will play an important role in the dynamical analysis of Chapter 4. The accuracy of these results have been verified through comparison with both experimental findings and other simulations found in the literature [B1, G1, H1, N1, V1].

To summarize, the curves of Figures 3.1.1, 3.2.1, and 3.2.2 represent the calculated time courses of the membrane potential, the controlling parameters, and the underlying potassium and sodium current and conductance changes at $T = 6.3$ °C (adapted from Hodgkin and Huxley's original experiment [H2, N1]).

# Chapter 4. The Dynamical Description: The Hodgkin-Huxley Membrane as a Bang-Bang Controller

## 4.1. Overview

In what follows, efforts will be directed at investigating the correction and switching properties of the Hodgkin-Huxley membrane as it undergoes changes in permeability. Perturbation equations for subthreshold oscillations will be introduced to describe the membrane permeability return-to-rest following an action potential. In particular, perturbations for a reduced, two-variable model of the Hodgkin-Huxley system will be examined. Nonlinear stability and transient analysis methods will follow and be used in developing the control policy.

## 4.2. Methods

The dynamical analysis will begin by introducing a reduced, linear, two-variable model of the original Hodgkin-Huxley system [H2]. This two-variable model is taken from a paper published by Rinzel [R2], and is the result of a formally developed systematic scaling treatment. This development is based on earlier work (although less formal) reported by Sabah [S1], and accounts for subthreshold oscillatory behavior of the membrane during changes in permeability.

In proceeding to apply Liapunov's first and second methods, the objective will be to provide a rigorous description of the governing switching mechanisms which act to stabilize membrane permeability. This method of derivation will include the development of a bang-bang control policy. The motivation and value for development of this policy is to examine the underlying mechanisms which act to optimize the performance of the membrane regulation properties.

## 4.3. Stability and Control in the Hodgkin-Huxley Membrane

The Hodgkin-Huxley system (2.1-1) through (2.1-4) may be approximated by a two-dimensional system that is amenable to a phase-plane analysis [H3]. The stiff nature of the Hodgkin-Huxley differential system suggests separating the kinetic processes into slow and fast modes. Accordingly, subthreshold oscillations during changes in membrane permeability may be approximated by considering Rinzel's two-state variable model [R2] in $(E, n)^{\ddagger}$. Rinzel employs a scaling treatment by the Jacobian matrix **J** based on a membrane resting potential of $E_o = -63.1\ mV$ [H1, H2] when $T = 6.3\ ^{\circ}C$

$$\mathbf{J} = \begin{bmatrix} -\partial\zeta/\partial E & -\partial\zeta/\partial m & -\partial\zeta/\partial n & -\partial\zeta/\partial h \\ f_2 m_\infty & -f_2 & 0 & 0 \\ f_6 n_\infty & 0 & -f_6 & 0 \\ f_4 h_\infty & 0 & 0 & -f_4 \end{bmatrix} \qquad (4.3\text{-}1)$$

where $\zeta \Rightarrow \zeta(E, m_\infty, n_\infty, h_\infty) + I_s(t) = 0$, $m_\infty \Rightarrow m_\infty(E) = \alpha_m(E)/\alpha_m(E) + \beta_m(E)$, and similarly for $n_\infty$ and $h_\infty$.

This scaling by **J** suggests an approximation on an immediate time-scale in which small changes in $m$-sodium (i.e., $\delta m$) relaxes infinitely *fast* and $\delta h$ is held at a nearly constant value. Denoting, then, the approximate variables $(\delta E, \delta n)$ and considering the value of $\delta n$ to being approximately zero for the decay of small displacements from a state of rest, the reduced, linear, two-variable Hodgkin-Huxley system for subthreshold oscillations may be expressed as:

‡ Incidentally, this two-variable system has been experimentally verified from data on small-signal responses in a squid axon [R2].

$$\delta\dot{E} = -0.24569\delta E - 1.00725\delta n$$

(4.3-2)

$$\delta\dot{n} = 0.154405\delta E - 0.1831975\delta n$$

or, in keeping consistent with the state variable assignment of Chapter 3

$$\delta\dot{x}_1 = -0.24569\delta x_1 - 1.00725\delta x_4$$

(4.3-3)

$$\delta\dot{x}_4 = 0.154405\delta x_1 - 0.1831975\delta x_4$$

## <u>4.3.1. Liapunov Function Generation Case 1: No Control Input ($u = 0$)</u>

From the discussion in Section 4.3, the linearized model expressed in (4.3.3) reflects a linear autonomous system $d\mathbf{x}/dt = \mathbf{A}\mathbf{x}$, where

$$\mathbf{A} = \begin{bmatrix} -0.24569 & -1.0072 \\ 0.154405 & -0.1831975 \end{bmatrix} \qquad (4.3\text{-}4)$$

with the understanding that the membrane equilibrium state is marginally stable due to the complex eigenvalues of **A** (with negative real parts). Following *Aizerman's method* [M4] in forming a Liapunov function $V(\mathbf{x})$, select a quadratic form which is positive-definite (p.d.), containing the perturbations $\delta\mathbf{x}$ such that

$$V(\delta\mathbf{x}) = \delta\mathbf{x}^T\mathbf{Q}\delta\mathbf{x} \qquad (4.3\text{-}5)$$

where **Q** is a real and symmetric matrix for unspecified $q_{ij} = q_{ji}$. It follows that $\dot{V}(\delta\mathbf{x}) = \delta\dot{\mathbf{x}}^T\mathbf{Q}\delta\mathbf{x} + \delta\mathbf{x}^T\mathbf{Q}\delta\dot{\mathbf{x}}$, or

$$\dot{V}(\delta\mathbf{x}) = -\delta\mathbf{x}^T\mathbf{P}\delta\mathbf{x} \tag{4.3-6}$$

where $-\mathbf{P} = \mathbf{A}^T\mathbf{Q} + \mathbf{Q}\mathbf{A}$ and $\dot{V}(\delta\mathbf{x})$ is negative-definite (n.d.), from which the elements of $\mathbf{Q}$ ($q_{ij}$) may be specified.

Proceeding, note that the p.d. quadratic form (4.3-5) will yield

$$V(\delta\mathbf{x}) = q_{11}\delta x_1^{\,2} + 2q_{12}\delta x_1\delta x_4 + q_{22}\delta x_4^{\,2} \tag{4.3-7}$$

Now, it follows from (4.3-6) and $-\mathbf{P} = \mathbf{A}^T\mathbf{Q} + \mathbf{Q}\mathbf{A}$ that $V(\delta\mathbf{x}) = \delta\mathbf{x}^T\{\mathbf{A}^T\mathbf{Q} + \mathbf{Q}\mathbf{A}\}\delta\mathbf{x}$, from which

$$\begin{aligned}\dot{V}(\delta\mathbf{x}) = (-0.49138q_{11} + 0.30881q_{12})\delta x_1^{\,2} + (-2.0145q_{11} - 0.85778q_{12} + 0.30881q_{22})\delta x_1\delta x_4 \\ + (2.0145q_{12} - 0.366395q_{22})\delta x_4^{\,2}\end{aligned} \tag{4.3-8}$$

In selecting a p.d. $\mathbf{P}$ (so that $\mathbf{Q}$ is p.d. for the stable subthreshold oscillations of $\mathbf{A}$) the identity matrix $\mathbf{I}$ is incorporated as a first try. For $\mathbf{P} = \mathbf{I}$, real $q_{ij}$, is specified, or $q_{11} = 1.49952$, $q_{12} = -0.85220$ and $q_{22} = 7.41484$. Substituting each $q_{ij}$ into (4.3-7) gives

$$V(\delta\mathbf{x}) = 1.49952\delta x_1^{\,2} - 1.7044\delta x_1\delta x_4 + 7.41484\delta x_4^{\,2} \tag{4.3-9}$$

The quadratic form (4.3-5) — and hence, the Liapunov function $V(\delta\mathbf{x})$ of (4.3-7) — is indeed p.d. since $q_{11} > 0$ and $4q_{11}q_{22} - q_{12}^{\,2} > 0$ [B4]. Solving for $\dot{V}(\mathbf{x})$ along a solution curve of $\mathbf{x}(t)$ for the *nonlinear* system is accomplished by [A2]

$$\dot{V}(\mathbf{x}) = \frac{\partial V(\delta\mathbf{x})}{\partial \delta x_1}\dot{x}_1 + \frac{\partial V(\delta\mathbf{x})}{\partial \delta x_4}\dot{x}_4 \tag{4.3-10}$$

In taking the derivatives, the $\delta\mathbf{x}$ notation is henceforth disused. Thus

$$\begin{aligned}\dot{V}(\mathbf{x}) &= (3x_1 - 1.7x_4)\dot{x}_1 + (-1.7x_1 + 14.8x_4)\dot{x}_4 \\ &= 3x_1\dot{x}_1 - 1.7\dot{x}_1x_4 - 1.7x_1\dot{x}_4 + 14.8x_4\dot{x}_4\end{aligned} \qquad (4.3\text{-}11)$$

Upon substitution of the nonlinear system expressions (3.1-5*a*) and (3.1-5*d*) into (4.3-11) (for $u = 0$)

$$\begin{aligned}\dot{V}(\mathbf{x}) &= 3x_1[(k_2 - k_1x_1)x_2^3x_3 + (k_4 - k_3x_1)x_4^4 - k_5x_1 + k_6] \\ &\quad - 1.7[(k_2 - k_1x_1)x_2^3x_3 + (k_4 - k_3x_1)x_4^4 - k_5x_1 + k_6]x_4 \\ &\quad - 1.7x_1[f_5(x_1) - f_6(x_1)x_4] + 14.8x_4[f_5(x_1) - f_6(x_1)x_4]\end{aligned} \qquad (4.3\text{-}12)$$

but note that this expression contains terms in $x_2$ and $x_3$. Therefore, it is necessary to express these two variables in terms of the membrane potential $x_1$. This may be accomplished by recalling Equations (2.2-1) and (2.2-2)

$$I_{\mathrm{Na}} = m^3h\bar{g}_{\mathrm{Na}}(E - E_{\mathrm{Na}}) \qquad (2.2\text{-}1)$$

$$I_{\mathrm{K}} = n^4\bar{g}_{\mathrm{K}}(E - E_{\mathrm{K}}) \qquad (2.2\text{-}2)$$

and as defined in Chapter 3, $(E, m, h, n) = (x_1, x_2, x_3, x_4)$. Then from (2.2-1), $m^3h = I_{\mathrm{Na}} / \bar{g}_{\mathrm{Na}}(E - E_{\mathrm{Na}})$ or $I_{\mathrm{Na}} / \bar{g}_{\mathrm{Na}}(x_1 - E_{\mathrm{Na}}) = F(x_1)$. A similar expression can be formed from (2.2-2) and is defined here as $G(x_1)$. Substituting for $F(x_1)$ and $G(x_1)$

$$\begin{aligned}\dot{V}(\mathbf{x}) &= 3x_1[(k_2 - k_1x_1)F(x_1) + (k_4 - k_3x_1)G(x_1) - k_5x_1 + k_6] \\ &\quad - 1.7[(k_2 - k_1x_1)F(x_1) + (k_4 - k_3x_1)G(x_1) - k_5x_1 + k_6]x_4 \\ &\quad - 1.7x_1[f_5(x_1) - f_6(x_1)x_4] + 14.8x_4[f_5(x_1) - f_6(x_1)x_4]\end{aligned} \qquad (4.3\text{-}13)$$

Upon expansion of equation (4.3-13) and the collection of like terms, $\dot{V}(\mathbf{x})$ may be written in the following, compact form:

$$\dot{V}(\mathbf{x}) = -3\mathrm{K}(x_1)x_1 + 1.7[\mathrm{K}(x_1) + \Phi(x_1)]x_1x_4 - \Psi(x_1)x_4^2 \qquad (4.3\text{-}14)$$

where $\mathrm{K}(x_1) = k_1F(x_1) - k_2F(x_1)/x_1 + k_3G(x_1) - k_4G(x_1)/x_1 + k_5 - k_6/x_1 + 0.567f_5(x_1)/x_1$, $\Phi(x_1) = 8.14f_5(x_1)/x_1 + 1.7f_6(x_1)$, and $\Psi(x_1) = 14.8f_6(x_1)$. Effectuating a n.d. condition on $\dot{V}(\mathbf{x})$ requires that $\mathrm{K}(x_1) > 0$ and $\Psi(x_1) > 0 \ \forall \ x_1$ and that the

$$\det \begin{bmatrix} 3\mathrm{K}(x_1) & -0.85[\mathrm{K}(x_1) + \Phi(x_1)] \\ -0.85[\mathrm{K}(x_1) + \Phi(x_1)] & \Psi(x_1) \end{bmatrix} > 0$$

from which the range of permissible parameters for asymptotic stability are

$$-\Phi + 2.08\Psi - \Omega < \mathrm{K}(x_1) < -\Phi + 2.08\Psi + \Omega \qquad (4.3\text{-}15)$$

where $\Omega = \pm 0.408(\Psi)^{0.5}(-25\Phi + 26\Psi)^{0.5}$.

It has been numerically evaluated that $\mathrm{K}(x_1) > 0$ for three ranges of the membrane potential $x_1$. Proceeding through time, sequentially (i.e., from the initial value of the resting potential through the afterpotential phases), the three ranges of $x_1$ are: (*i*) $-63.1 \le x_1 < 22.8$ *mV* (resting through the excitation inflection point); (*ii*) $22.8 < x_1 < -19.4$ *mV* (plateau through the permeability transient); and (*iii*) $-47.1 < x_1 \le -65.3)$ *mV*. Furthermore, it was confirmed that $\Psi(x_1) > 0 \ \forall \ x_1$.

## 4.3.2. Systems Plots and Physical Interpretation of Results

The curve illustrated in Figure 4.3.2.1 shows $V(\mathbf{x})$ for the membrane as it undergoes changes in permeability during a complete action potential. There are three ranges of $x_1$ for which $V(\mathbf{x})$ is a Liapunov function (i.e., where $V$ is p.d. and $\dot{V}$ is n.d.). That is, where $K(x_1) > 0$ and $\Psi(x_1) > 0 \; \forall \; x_1$ (see Section 4.3.1). The regions of $V(\mathbf{x})$ where these latter conditions are satisfied are labeled **I, II** and **III**.

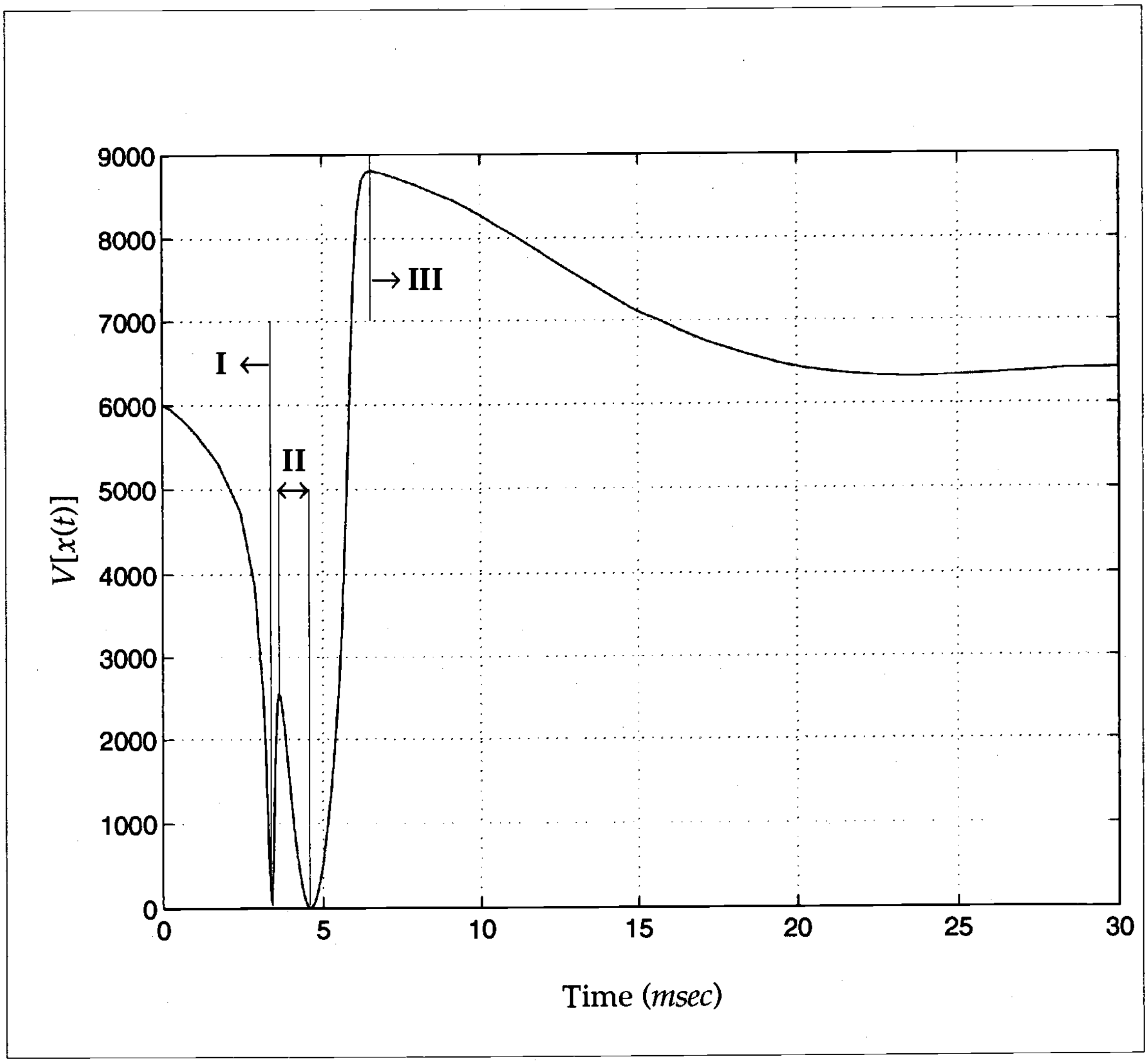

Figure 4.3.2.1. Liapunov function for the membrane (Regions **I, II,** and **III**).

A physical interpretation for each of these regions is given as follows: When the membrane is perturbed from rest, ionic potassium begins to decrease at a rate nearly equal to that of increasing sodium (Region I). The membrane then rapidly undergoes an unstable depolarization (fast mode) at which time it becomes highly permeable to *m*-sodium (Region II). Subsequent to the plateau in $x_1$, a brief duration of time elapses before $x_4$ becomes large enough to contribute to the potassium current flow needed to stabilize the membrane. When the membrane becomes predominantly permeable to *n*-potassium, hyperpolarization has occurred. This mode of regeneration is governed by $x_4$ and ultimately relaxes the Liapunov function (Region III). Figure 4.3.2.2 shows the positively damped permeability perturbations moving into regions bounded by hypersurfaces of constant $V$. Curves of constant $\dot{V}$ define the membrane attraction domain.

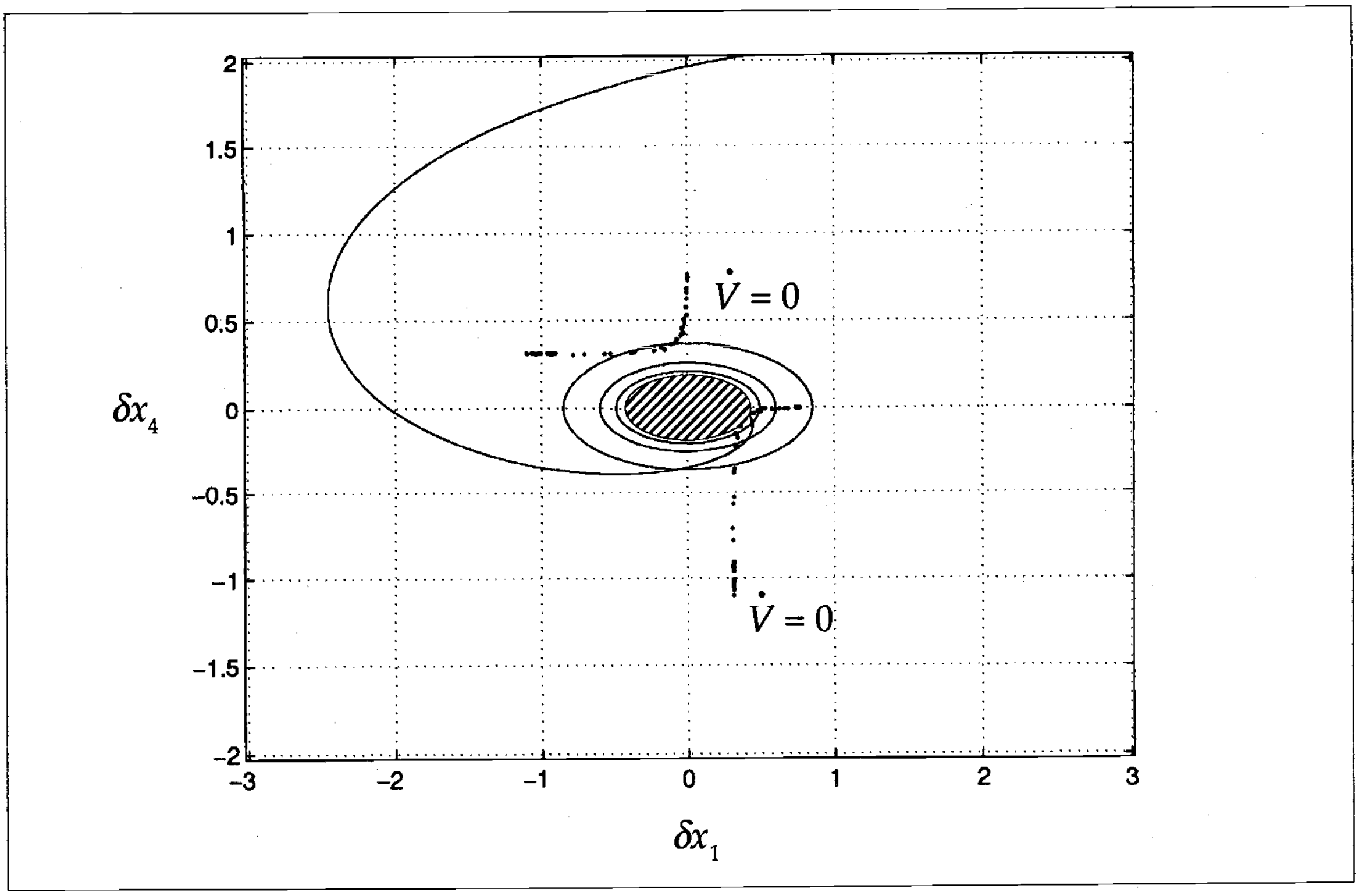

Figure 4.3.2.2. Contours of constant $V$ and attraction domain shown by constant $\dot{V}$.

That is, the curves of $\dot{V} = 0$ tangent to the cross-hatched elliptical region and the outermost ellipse define the Liapunov function as verified by (4.3-9).

The curve in Figure 4.3.2.3 illustrates the state-plane projected trajectories of the nonlinear membrane model.

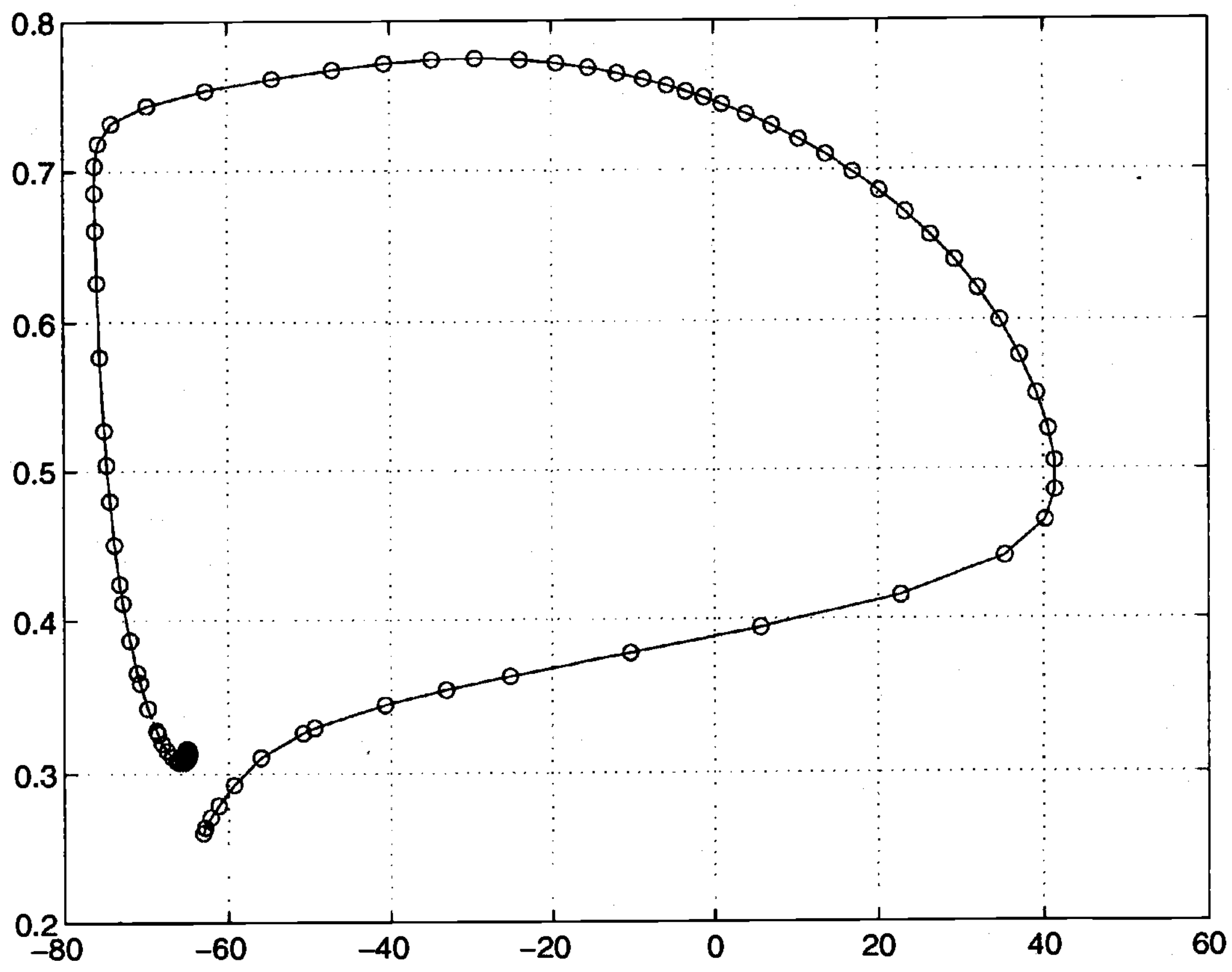

Figure 4.3.2.3. Membrane model state-plane trajectory during permeability changes.

As the state trajectory shows, the actual region of attraction to the equilibrium state is larger than that guaranteed by the given Liapunov function.

### 4.3.3. Case 2: Synthesis of a Control Policy ($u \neq 0$)

The objective is to synthesize a control policy to drive the membrane state $\delta\mathbf{x}(t)$ to equilibrium by minimizing the function $\dot{V}(\mathbf{x}, \mathbf{u})$ [M4] — given $V(\mathbf{x})$ — where $\mathbf{u} \in \mathfrak{R}^m$ is a control vector. This policy will be developed with the knowledge that the unforced system of (4.3-3) is asymptotically stable at rest with assumed $\dot{V}(\mathbf{x}) < 0$. Consider, then, control $\mathbf{u}$ added to the assumed constant current stimulus $I_s(t)$. The objective of this control is to *maximize* $-\dot{V}(\mathbf{x}, \mathbf{u})$, where

$$\dot{V}(\mathbf{x}, \mathbf{u}) = -\mathbf{x}^T\mathbf{P}\mathbf{x} + 2\mathbf{u}^T\mathbf{B}^T\mathbf{Q}\mathbf{x} \qquad (4.3\text{-}16)$$

where the vector $\mathbf{B}^T = [1 \quad 0]$ and hence, $2\mathbf{u}^T\mathbf{B}^T\mathbf{Q}\mathbf{x} = 2\mathbf{u}(q_{11}x_1 - q_{12}x_4)$. Numerical values for $q_{11}$ and $q_{12}$ were given in Section 4.3.1. In order to minimize (4.3-16) — thereby maximizing $-\dot{V}(\mathbf{x}, \mathbf{u})$ — then the form of the control is an adaptive one; specifically, a bang-bang process. If the scalar control $u$ is constrained such that $|u| \leq U$, the bang-bang control will take the form of

$$u = -U \operatorname{sgn} S(\mathbf{x}) \qquad (4.3\text{-}17)$$

where $S(\mathbf{x})$ is the so-called switching function defined as

$$S(\mathbf{x}) = (\mathbf{B}^T\mathbf{Q}\mathbf{x}) \qquad (4.3\text{-}18)$$

Based on the linearized Hodgkin-Huxley system (4.3-3) with added control $u$, the numerical values of $q_{11}$, $q_{12}$ and the analysis performed in § 4.3.1, (4.3-16) becomes

$$\dot{V}(\mathbf{x}, \mathbf{u}) = -3\mathrm{K}(x_1)x_1^2 + \Lambda(x_1)x_1x_4 - \Psi(x_1)x_4^2 + (3x_1 - 1.7x_4)u \qquad (4.3\text{-}19)$$

where $\Lambda(x_1) = 1.7[\mathrm{K}(x_1) + \Phi(x_1)]$. Furthermore, the switching function is identified as $S(\mathbf{x}) = 3x_1 - 1.7x_4$. To this end, the bang-bang control policy for membrane permeability changes is written as

$$u = -U \operatorname{sgn} \{3x_1 - 1.7x_4\} \tag{4.3-20}$$

for the objective of minimizing $-\dot{V}(\mathbf{x}, \mathbf{u})$.

## 4.4. Suggested Method for Introducing the Additive Control

In suggesting a means for introducing the additive membrane control, consider the ionic potassium current term of (2.2-2)

$$I_K = n^4 \bar{g}_K (E - E_K) \tag{2.2-2}$$

where the maximum potassium conductance $\bar{g}_K$ is known to act as a bifurcation parameter [H3]. It is therefore suggested that the membrane control policy of (4.3-20) be realized according to this effect.

As this parameter is reduced from its standard value of 36 $mmho/cm^2$, the Hodgkin-Huxley membrane model [H2] will actually sustain repetitive firing. This seemingly unnatural phenomenon [B3, C1, M7, S1] was discussed in Chapter 2 for the case of the Bonhoeffer-Van der Pol system (2.2-5). The Hodgkin-Huxley membrane will sustain repetitive firing for $\bar{g}_K = 18$ $mmho/cm^2$. Notably, Hopf bifurcations will occur close to $\bar{g}_K = 19.76$ $mmho/cm^2$, whereas stable equilibrium occurs at $\bar{g}_K = 24$ $mmho/cm^2$.

This suggests the control is a constrained one. But since the policy is not explicit in $\bar{g}_K$, the additive control must be introduced through regulating the flow of the potassium current through the branch containing the *time-varying* conductance $g_K$ (i.e., in controlling the flow of $I_K$, we implicitly are regulating $\bar{g}_K$). In doing so, switching limits are placed on the potassium controlling parameter $x_4$, thereby maintaining the bifurcation parameter within a stable range. Indeed, the optimal control policy ensures this by rapidly driving the state to the resting potential by maximizing $-\dot{V}(\mathbf{x}, \mathbf{u})$. Physically this means that the concentration of potassium is rapidly increased until the resting potential state has been reached.

From an electrical engineering viewpoint, this control could be realized through the pulse network of Figure 2.2.2. This circuit effectively simulates the excitability and threshold phenomena of the Hodgkin-Huxley nerve membrane as described by the Bonhoeffer-Van der Pol system of (2.2-5). Firing of the diode device is readily controllable through the resistance region of $D_T$, the required parameter values for *R*, *L*, and *C*, and an appropriate trigger pulse [M2] to drive the controlling tunneling current $I_T$.

## 4.5 Conclusion

A bang-bang control policy describing stability and permeability changes in a Hodgkin-Huxley membrane model [H2] was developed. This foregoing policy demonstrates how the inherent mechanisms of a membrane act to drive its permeability from depolarization to the recovery and resting potential states‡.

It was discussed that these permeability changes in state occur along stable trajectories as governed by a switching function $S(\mathbf{x})$. It was shown that this switching function is dependent on the membrane action potential and the potassium controlling parameter. This dependence on the latter suggests that the underlying mechanism governing the maximization of $-\dot{V}(\mathbf{x}, \mathbf{u})$ lies in where the concentration of potassium rapidly increases until the resting potential has been reached. This increase indicates, physiologically, the rapid opening of the potassium channel. From the aspects of the Hodgkin and Huxley model [H2], this is the so-called hyperpolarization process.

‡ Interestingly, this appears to tie-in with Bernstein's 1902 hypothesis, which suggests that the membrane in a state of rest is semipermeable only to potassium.

The optimal policy ensures that this process will take place rapidly (on the order of milliseconds) by maximizing $-\dot{V}(\mathbf{x}, \mathbf{u})$ for the specified Liapunov function $V(\mathbf{x})$. The maximum potassium conductance $\bar{g}_K$ was defined as a bifurcation parameter which is controlled through regulation of the potassium current within the membrane. It was suggested that this control be realized through an approximate second-order network described by a set of Bonhoeffer-Van der Pol equations.

## 4.6. Suggestions for Future Directions

It would appear that permeability correction arises rather naturally in the membrane system. The task of manipulating system parameters in the laboratory through a carefully modeled admissible control should be left to the biophysicist or neural physiologist. It is hoped that the analysis presented here provides a base in carrying out such systematic experimental planning and data collection. For the computer engineer, this research might possibly provide a base for developing a rather simple, but realistic architecture which mimics the propagation of an action potential from one cell to another.

# Appendix

# Appendix. Computer Code used in Simulating the Hodgkin–Huxley Equations of Ionic Hypothesis

In reference to Sections 3.1 and 3.2, the plots illustrated in Figures 3.1.1, 3.2.1 and 3.2.2 are provided as a complete summary of all simulation results of the original, Hodgkin-Huxley model. These plots represent solutions of the membrane permeability changes (surrounding the squid axon ) during a propagated action potential.

The Hodgkin-Huxley model is represented by state equations (3.1-5*a*) through (3.1-5*d*) (as seen in Section 3.1). These equations were integrated on an HP Apollo 400 Series UNIX platform using a fourth-order Runge-Kutta algorithm. The following pages contain the entire set of M-files used in this simulation.

In keeping consistent with Hodgkin and Huxley's original experiment (in addition with the current literature), standard values of the membrane parameters used in the simulation have included: $E_{Na} = 50\ mV$; $\bar{g}_{Na} = 120\ mmho/cm^2$; $E_K = -77\ mV$; $\bar{g}_K = 36\ mmho/cm^2$; $E_L = -54.387\ mV$; and $\bar{g}_L = 0.3\ mmho/cm^2$. The membrane capacitance was set to $C_M = 1\ \mu F/cm^2$.

The resting potential was set at a value of $E_o = -63.1\ mV$, where as the initial controlling parameters $m_o$, $h_o$, and $n_o$ were set to values of 0.0355402, 0.705487, and 0.260777, respectively. The applied stimulus current was set to $I_s(t) = -0.2\ mA$.

Dec 15 13:57 /amaterasu/sd16d/1abf94/melendyr/BH.m

```
function xdot = BH(t,x)
%
%
%
global Is Cm gNa gK gL ENa EK EL alphaM BetaM alphaH BetaH alphaN BetaN
%
global ks k1 k2 k3 k4 k5 k6 f1 f2 f3 f4 f5 f6
%
%
%
%
%
%       Is = -(sign(t)-sign(t-1));
Is = -0.2;
%       Is = -0.2*stepfun(1:.1:5,2)+0.2*stepfun(1:.1:5,2.2);
%
Cm = 1; gNa = 120; gK = 36; gL = 0.3;
%
ENa = 50; EK = -77; EL = -54.387;
%
ks = Is/Cm; k1 = gNa/Cm; k2 = (gNa*ENa)/Cm; k3 = gK/Cm; k4 = (gK*EK)/Cm;
%
k5 = gL/Cm; k6 = (gL*EL)/Cm;
%
%
%
%
%
%       *****************************************************************
%       *                                                               *
%       *       Type-"M" Sodium (Na+) gate:  alpha_M and Beta_M         *
%       *       rate-constants (expressions for):                       *
%       *                                                               *
%       *****************************************************************
%
alphaM = 0.1*(x(1) + 40)/(1 - exp(-(x(1) + 40)/10));
%
BetaM = 0.108*exp(-x(1)/18);
%
f1 = alphaM; f2 = (alphaM + BetaM);
%
%
%
%
%
%       *****************************************************************
%       *                                                               *
%       *       Type-"H" Sodium (Na+) gate:  alpha_H and Beta_H         *
%       *       rate-constants (expressions for):                       *
%       *                                                               *
%       *****************************************************************
%
alphaH = 0.0027*exp(-x(1)/20);
%
BetaH = 1/(1 + exp(-(x(1) + 35)/10));
%
f3 = alphaH; f4 = (alphaH + BetaH);
%
%
%
%
%
%
%
%       *****************************************************************
%       *                                                               *
%       *       Type-"N" Potassium (K+) gate:  alpha_N and Beta_M       *
%       *       rate-constants (expressions for):                       *
%       *                                                               *
%       *****************************************************************
%
alphaN = 0.01*(x(1) + 55)/(1 - exp(-(x(1) + 55)/10));
%
BetaN = 0.0555*exp(-x(1)/80);
%
f5 = alphaN; f6 = (alphaN + BetaN);
%
%
%
%
%
%       *****************************************************************
%       *                                                               *
%       *               Hodgkin-Huxley Differential Equations:          *
%       *                                                               *
%       *****************************************************************
%
%
xdot(1) = ks + (k2-k1*x(1))*x(3)*(x(2)^3) + (k4-k3*x(1))*(x(4)^4) -k5*x(1) + k6;
xdot(2) = f1 - f2*x(2);
xdot(3) = f3 - f4*x(3);
xdot(4) = f5 - f6*x(4);
%       xdot = xdot';
```

```
Dec 15 13:58 /amaterasu/sdl6d/labf94/melendyr/BHexe.m

%       ************************************************************
%       *                                                          *
%       *     Set up the rate-constant differential equations      *
%       *     from "BH.m" for a numerical solution using ode45.    *
%       *                                                          *
%       ************************************************************
%
%
global Is Cm gNa gK gL ENa EK EL alphaM BetaM alphaH BetaH alphaN BetaN
%
global ks k1 k2 k3 k4 k5 k6 f1 f2 f3 f4 f5 f6
%
%
t0 = 0;         % The starting time for the integration (t_initial).
%
tf = 30;
%
tfinal = 30;    % The ending time for the integration (t_final).
%
tspan = [t0 tfinal];
%
x0 = [-63.1 0.0355402 0.705487 0.260777]';
%
tol = 1.e-3;    % The desired accuracy of the solution.
%
%
%
%       ************************************************************
%       *                                                          *
%       *                     Invoke ode45:                        *
%       *                                                          *
%       ************************************************************
%
%
[t,x] = ode45('BH',t0,tf,x0,tol);
%
%       [t,x] = ode45('BH',tspan,x0);
%
% Calls the M-file "BH.m" and solves for the probabilities m, h, n, and
% the membrane voltage E.
%
%
%
%       ************************************************************
%       *                                                          *
%       *     Plot Na+ and K+ controlling parameters m, h, and n;  *
%       *     Plot the membrane potential E.                       *
%       *                                                          *
%       ************************************************************
%
%
figure(1); subplot(2,1,1); plot(t,x(:,1),'r');
%
%       xlabel('Time (msec)'); ylabel('E (mV)');
%
%
%
%
figure(1); subplot(2,1,2); plot(t,x(:,2),'g'); hold on;
%
plot(t,x(:,3),'b'); plot(t,x(:,4),'m');
%
%       xlabel('Time (msec)'); ylabel('Probability');
%
%
%
%
%
%       ************************************************************
%       *                                                          *
%       *     Extract and plot the Na+ and K+ ionic currents;      *
%       *     Likewise, extract and plot the Na+ and K+ total      *
%       *     conductance changes from INa and IK, respectively.   *
%       *                                                          *
%       ************************************************************
%
%
INa = (x(:,2).^3).*x(:,3).*gNa.*(x(:,1) - ENa); GNa = INa./(x(:,1) - ENa);
%
IK = (x(:,4).^4).*gK.*(x(:,1) - EK); GK = IK./(x(:,1) - EK);
%
%
%
%
figure(2); subplot(2,1,1), plot(t,INa,'g'); hold on; plot(t,IK,'m');
%
%       xlabel('Time (msec)'); ylabel('mA/cm^2');
%
%title('Fig. 2.  Sodium and Potassium Current Changes (T = 6.3 degrees C)');
%
%
%
%
figure(3); plot(t,GNa,'g'); hold on; plot(t,GK,'m');
%
%       xlabel('Time (msec)'); ylabel('mmho/cm^2');
%
%title('Fig. 3.  Sodium and Potassium Conductance Changes (T = 6.3 dC)');
%
IL = gL.*(x(:,1) - EL); Ii = (INa + IK + IL); CEdot = -(Is + INa + IK + IL)*Cm;
%
%
figure(2); subplot(2,1,2), plot(t,CEdot);
%
%       xlabel('Time (msec)'); ylabel('mA/cm^2');
%
%       title('Fig. 2.  Capacitive Component (T = 6.3 degrees C)');
%
x1dot = ks+(k2-k1.*x(:,1)).*x(:,3).*(x(:,2).^3)+(k4-k3.*x(:,1)).*(x(:,4).^4)-k5.
%
%
%
%
%
%
%
%
%       ******************************************************
%       *                                                    *
%       *     Implement results from the dynamical analysis  *
%       *     of Chapter 4:  Liapunov's Direct Method and    *
%       *     Bang-Bang Control results                      *
%       *                                                    *
%       ******************************************************
%
%
%       ****n.d. condition where superdiagonal functions > 0****
%
%
```

```
F1 = INa./gNa.*(x(:,1) - ENa);
G1 = IK./gK.*(x(:,1) - EK);
a = k1*F1; b = k2*F1./x(:,1); c = k3*G1; d = k4*G1./x(:,1);
e = k5; f = k6./x(:,1);
alphaN = 0.01*(x(:,1) + 55)./(1 - exp(-(x(:,1) + 55)/10));
BetaN = 0.0555*exp(-x(:,1)/80);
f6 = (alphaN + BetaN);
%
%
K = a - b + c - d + e - f + 0.567*f6./x(:,1);
%
%
%       ****Find those positive x-values which render K(x1) > 0****
%
%
I = find(K>0); x(I);
%
%
%
%
figure(4); plot(x(:,1),x(:,4),'bo'); hold on; plot(x(:,1),x(:,4),'m-');
%
%       xlabel('Membrane Potential (mV)');
%       ylabel('Potassium Controlling Parameter');
%       title('Fig. 4.  Phase-Plane Trajectory for K+ and Membrane Potential');
%
%
V = 1.49952.*(x(:,1).^2)-1.7044.*(x(:,1).*x(:,4))+7.41484.*(x(:,4).^2);
%
figure(5); plot(t,V,'m');
%
%       xlabel('Time (msec)');
%       ylabel('V[x(t)]');
%       title('Fig. 5.  Liapunov Function of the Membrane');
%
%
figure(6); plot3(t, x(:,1), K, 'bo'); hold on;
figure(6); plot3(t, x(:,1), K, 'm-'); grid;
%
%
figure(7); plot3(x(:,1), x(:,4), K, 'bo'); hold on;
figure(7); plot3(x(:,1), x(:,4), K, 'm-'); grid;
%
%
f5 = alphaN;
Phi = 8.14*f5 + 1.7*f6;
Psi = 14.8*f6;
Omega = 0.408*((Psi).^(0.5)).*(-25*Phi + 26*Psi).^(0.5);
KL = -Phi + 2.08*Psi - Omega;
KU = -Phi + 2.08*Psi + Omega;
figure(8); plot(x(:,1), KL, 'c'); hold on;
figure(8); plot(x(:,1), KU, 'r'); grid;
%
%
figure(9); plot3(t, x(:,1), V, 'bo'); hold on;
figure(9); plot3(t, x(:,1), V, 'm-'); grid;
%
%
figure(10); plot3(x(:,4), x(:,1), V, 'bo'); hold on;
figure(10); plot3(x(:,4), x(:,1), V, 'm-'); grid;
%
%
figure(11); plot3(x(:,1), x(:,4), V, 'bo'); hold on;
figure(11); plot3(x(:,1), x(:,4), V, 'm-'); grid;
```